\documentstyle[preprint,revtex]{aps}
\begin{document}
\draft
\begin{title}
CHARACTERISTIC POTENTIALS FOR MESOSCOPIC
\end{title}
\begin{title}
RINGS THREADED BY AN AHARONOV-BOHM FLUX
\end{title}
\author{M. B\"{u}ttiker}
\begin{instit}
IBM T. J. Watson Res. Ctr., P. O. Box 218,
Yorktown Heights, N. Y. 10598
\end{instit}
\receipt{Today}
\begin{abstract}
Electro-static potentials for samples with the topology of a
ring and penetrated by an Aharonov-Bohm flux are discussed.
The sensitivity of the electron-density distribution
to small variations in the flux generates an effective
electro-static potential which is itself a periodic function
of flux. We investigate a simple model in which the flux
sensitive potential
leads to a persistent current which is enhanced compared to
that of a loop of non-interacting electrons.
For sample geometries with contacts the sensitivity of the
electro-static potential to flux leads to a flux-induced
capacitance.
This capacitance gives the variation in charge due to an
increment in flux.
The flux-induced capacitance is contrasted with the
electro-chemical capacitance which gives the variation
in charge due to an increment in an electro-chemical potential.
The discussion is formulated in terms of characteristic functions
which give the variation of the electro-static potential in the
interior of the conductor due to an increment in the external
control parameters (flux, electro-chemical potentials).
\end{abstract}

\narrowtext

\section{Introduction}

In this work we discuss
the electro-static potential in the interior of small
mesoscopic conductors with the topology of a ring (see Fig. 1).
An Aharonov-Bohm flux (AB-flux) penetrates the hole of the loop.
Typically the electron-density distribution in such a loop
is non-uniform and is a sensitive function of the AB-flux.
The non-uniform density generates an electro-static
potential which is also a function of the flux.
We introduce a characteristic potential $v({\bf r})$
which gives the variation of the electro-static
potential in the interior of conductors
in response to an increment of the AB-flux.
We investigate a simple example of a one-dimensional loop
structure (see Fig. 2)
which permits a solution for the characteristic potential
$v({\bf r})$.
This example provides a demonstration that interaction
can enhance the persistent current above its value in
the absence of interactions.
The characteristic potential
$v({\bf r})$ is used to
define a flux-induced capacitance which
is the ratio of the charge increment on
the conductor divided by the increment in flux.
The flux-induced capacitance is an odd (periodic) function of flux.
For samples with contacts
we introduce a
characteristic potential
$u({\bf r})$ which gives the
variation of the
local potential inside the conductor in response to an
increment of an electro-chemical potential at a contact.
This characteristic function allows the evaluation
of the electro-chemical capacitance.
The electro-chemical capacitance is an even
periodic function of flux with
period $\Phi_{0}= hc/e$.
The electro-chemical capacitance can be measured
with the help of small time-dependent,
oscillating voltages applied to a capacitor.
The flux-induced capacitance
is measured by applying a small oscillatory flux superimposed
on a steady state AB-flux.

Recent work\cite{MBHTP} by
Thomas, Pr\^{e}tre and this author on the admittance of mesoscopic
capacitors
emphasized the need to distinguish between
electric and electro-chemical capacitances.
Capacitance coefficients are not
electro-static entities determined by the geometry of the sample alone
but are electro-chemical quantities
which depend on the properties of the conductor\cite{MBCAP}.
The non-geometrical contributions
to the capacitance coefficients arise due to the
fact that electric fields impinging on a conductor
are not screened immediately
at the surface of the conductor but penetrate over
a distance of a screening length
into the "bulk" of the conductor\cite{STERN}.
Clearly field penetration is
very important
for mesoscopic conductors\cite{MBHTP}
since one or more dimensions of such
a small system might in
fact be comparable to a screening length!
A dramatic demonstration of the
non-geometrical nature of
capacitances is given by Chen et al.\cite{WEI94} who
show that magnetic fields
can completely quench certain elements of a capacitance tensor.

In samples without transmission the
capacitance coefficients are an equilibrium phenomenon
which reflects properties of the ground state.
Another interesting ground state
property of a small mesoscopic ring is the
equilibrium persistent current.
The existence of such currents in mesoscopic
{\it disordered} normal rings was
predicted by this author in collaboration with
Imry and Landauer\cite{MBYIL}.
In earlier work persistent currents
were discussed for large molecules\cite{PAULI,LONDO},
Landau diamagnetism\cite{HUNDF},
and in connection with superconductivity\cite{BEYER,BLOCH}.
For many years
equilibrium currents in mesoscopic
conductors remained of interest to only
a small
community\cite{RLB85,MBSMA,MBK86,CHEUN,BOUCH}.
Experimental observation of such equilibrium currents in an
ensemble of rings by
Levy et al.\cite{LEVYD},
in single metallic loops by Chandrasekhar et al.\cite{CHAND},
and single semiconductor rings by Mailly et al.\cite{MAILL}
generated a considerable theoretical literature.
We can refer the reader here to only a few representative
contributions investigating the persistent current of
non-interacting\cite{AGI91,RIEDE,AKKER} and
interacting\cite{SCHMI,LOMAR,KOPIE,ARGAM,MUWEI} electrons.
Persistent currents occur
not only in isolated rings but also in rings
connected
via leads to electron reservoirs\cite{MBSMA,CHEUN,AKKER,MELLO,JAYAN}.
The experiment of Mailly et al.\cite{MAILL}
in fact measures persistent currents in
both closed and open rings of roughly the same amplitude.
Since both capacitances and persistent
currents reflect properties of the ground
state it is intriguing to explore possible
connections between the two phenomena.

\section{Self-Consistent Potential of a Normal Loop}

Consider the conductors in Fig. 1.
The conductor of Fig. 1a is via a lead connected to an electron
reservoir\cite{MBSMA,MELLO,JAYAN}.
It is otherwise in an electrically isolating environment:
There are no electric field lines penetrating the surface of a volume
which is at a sufficient distance from the loop and intersects the
reservoir far enough from the connection to the lead.
Typically conductors are not in such an electrically isolated
environment but couple via long range Coulomb forces to other
nearby metallic conductors.
Such a situation is investigated for the conductor of Fig. 1b.
Here the conductor of Fig. 1a is separated via an insulator from
a second conductor, a gate,
which is connected to an electron-reservoir
at electro-chemical potential $\mu_{2}$.
Again we will for clarity assume that there exists a volume which
encloses both conductors and a portion of the reservoirs
such that no electric-field lines penetrate its surface.
In both structures one of the conductors forms a loop
that is threaded by an AB-flux.
The equilibrium electro-static potential
$U(\mu_{1},\mu_{2},\Phi, {\bf r})$
for these conductors is
a function of the electro-chemical potentials,
the flux and a complicated function of position.
The flux-dependence of the potential is generated by the
electron-density of the loop which is flux dependent
whenever the loop is not rotationally invariant.
We are interested in the variation of the electro-static potential
under small changes of the external control parameters
$\mu_{1},\mu_{2}$ and $\Phi$.
Small increases
$d\mu_{1},
d\mu_{2},
d\Phi$ in the electro-chemical potentials and or the flux
will bring the conductor from the equilibrium
state with potential
$U(\mu_{1},\mu_{2},\Phi, {\bf r})$
to a
new equilibrium state with potential
$U(\mu_{1}+d\mu_1,\mu_{2}+d\mu_2, \Phi + d\Phi, {\bf r})$.
The difference $dU$ between these two equilibrium potentials
can be expanded in powers of the increment in electro-chemical potential
and the increment in flux. To linear order we have
\begin{eqnarray}
edU(\mu_{1}, \mu_{2},\Phi, {\bf r}) =
u_{1}({\bf r}) d\mu_{1}
+u_{2}({\bf r}) d\mu_{2}
+v({\bf r}) d\Phi.
\label{eq18}
\end{eqnarray}
Here
$u_{k}({\bf r}) =$
$edU(\mu_{1}, \mu_{2}, \Phi, {\bf r})/
d\mu_{k}|_{d\mu_{k} = 0},$ with $k = 1, 2$ and
$v({\bf r}) =
edU(\mu_{1}, \mu_{2}, \Phi, {\bf r})/
d\Phi|_{d\Phi= 0}$  are
the {\it characteristic}
{\it functions}\cite{MBCAP}.
We will use the characteristic functions
$u_{k}({\bf r}) $ to derive expressions for the
electro-chemical capacitance $C_{\mu}=dQ/d\mu.$
In analogy to the electro-chemical capacitance we also
derive with the help of the characteristic function $v$
a flux-induced capacitance $C_{\Phi} = dQ/d\Phi$.
For the conductor of Fig. 1a which is an electrically isolating
environment an increase in the electro-chemical potential of
the reservoir or an increase in flux cannot change the overall
charge. For this conductor both the electro-chemical
capacitance and the flux-induced capacitance vanish.
The conductor of Fig. 1a is thus of interest only as
a simple limiting case.

For the conductor of Fig. 2b,
the characteristic potentials
$u_{1}$ and $u_{2}$ have the following
interesting properties\cite{MBCAP}:

(1) For ${\bf r}$ deep in reservoir $k$
the local potential must follow the electro-chemical potential
of that reservoir and hence
$u_{k}({\bf r}) = 1.$

(2) A change in the electro-chemical potential of reservoir
$k$ cannot affect the potential deep inside any
other reservoir. Thus for ${\bf r}$ deep inside reservoir
$l \ne k$ the characteristic function must vanish,
$u_{k}({\bf r}) = 0.$

(3) If we change all electro-chemical potentials
simultaneously and by an equal amount
$d\mu_{k}=d\mu$
then we have only changed our (global)
energy scale. Hence at every space point ${\bf r}$ the
potential $U$ must also change by $d\mu$.
This implies that the sum of all characteristic functions
at every space point is equal to 1,
\begin{eqnarray}
\sum_{k} u_{k}({\bf r}) = 1 .
\label{eq19}
\end{eqnarray}

In contrast to the characteristic potentials $u_{k}$ which approach
1 in contact $k$, an increment in flux polarizes the sample
but does not affect the state of the system deep inside a reservoir.
Hence the characteristic function $v({\bf r})$ vanishes deep inside
the electron reservoirs.

To treat the Coulomb interactions it is useful to take the
density functional theory as a guide\cite{KOHNW,NLANG}.
A discussion of persistent currents based on this approach
is given in Ref. \cite{ARGAM}.
We are interested in
the magnitude of the variation of the
potential with flux.
To this extend we will discuss the closed
loop structure in Fig. 2.
If the loop is taken to be in an electrically isolating
environment
the electro-static
potential is a function of flux only.
(There is no dependence on a chemical potential).
The single particle
wave functions
of the ground state of the ring are thus determined
by a Hamiltonian
\begin{eqnarray}
H_{eff} = \frac{1}{2m}({\bf p}
- \hbar \frac{2\pi}{L} \frac{\Phi}{\Phi_0})^{2}
+eU_{eff}(\Phi, {\bf r}),
\label{eq03}
\end{eqnarray}
Here
$\Phi$ is the flux,
$\Phi_0 = hc/e$ is the single charge flux quantum, and
$L=2\pi R$ is the circumference of the loop.
The effective potential $eU_{eff}(\Phi, {\bf r})$
contains in addition to the electro-static potential
an exchange potential\cite{KOHNW,NLANG}.
The total kinetic energy of the electrons in the loop is
\begin{eqnarray}
E_{kin} = \sum E_{n}(\Phi, U(\Phi, {\bf r}))
- \int d^{3}{\bf r} n(\Phi, {\bf r})
eU_{eff}(\Phi, {\bf r}),
\label{ekin}
\end{eqnarray}
where
$E_{n}(\Phi, U(\Phi, {\bf r}))$
are the eigenvalues of Eq. (\ref{eq03}) and
$n(\Phi, {\bf r})$ is the electron-density.
The sum is over all occupied levels.
The persistent current of this loop is the flux-derivative
of the free energy.
At $kT = 0$ the free energy is equal to the total internal
energy $E_{int} = E_{kin}+E_{c}+E_{ex}$,
where $E_{c}$ is the Coulomb interaction energy and
$E_{ex}$ is the exchange energy
\cite{KOHNW,NLANG}.
At $kT = 0$ the persistent current of a closed loop is
determined by
\begin{eqnarray}
I_{eq}(\Phi) = -c dE_{int}/d\Phi .
\label{pers}
\end{eqnarray}
The persistent current does not in an explicit way depend
on the characteristic function $v(\Phi)$ introduced above.
This is a consequence of the fact that for potential
variations away from the true equilibrium potential
the internal energy is stationary\cite{NLANG}.
As remarked already in Ref. \cite{ARGAM},
the total flux derivative in Eq. (\ref{pers}) can be replaced
by a partial derivative taken at constant potential $U$.
Nevertheless, the potential variation with flux is important:
The spectrum of the interacting and the non-interacting system
are not the same.

For the open conductors of Fig. 1 it is
the grand canonical potential,
$\Omega = E_{int} - \sum_{k} N_{k}\mu_{k}$ that counts.
The persistent current in these structures is
\begin{eqnarray}
I_{eq}(\Phi) = -c d{\Omega}/d{\Phi}.
\label{per2}
\end{eqnarray}
The electro-chemical capacitance coefficients are
\begin{eqnarray}
C_{\mu,kl} (\Phi) = -e^{2} d^{2}\Omega/d\mu_{k}d\mu_{l}.
\label{cap1}
\end{eqnarray}
and the flux-induced capacitances are
\begin{eqnarray}
C_{\Phi,k} (\Phi) = -e d^{2}\Omega/d\Phi d\mu_{k}.
\label{capp}
\end{eqnarray}
{}From this definition of the flux-induced capacitance
another alternate interpretation of the second order
mixed derivatives of the grand canonical potential
becomes apparent.
Since
$I_{eq}(\Phi) = -c d\Omega/d\Phi$
the flux induced capacitance is related to a "conductance"
$G_{\Phi,k} = c C_{\Phi,k} = -ec (dI_{eq}/d\mu_{k})$
which is a measure of the sensitivity of the persistent
current to a small change in the electro-chemical potential.
The flux-induced capacitances are thus connected to the
gate voltage dependence of the persistent current.

We make a clear distinction between
{\it closed} loops with a fixed number of carriers and {\it open}
loops which are via leads connected to electron reservoirs.
Much of the mesoscopic literature attempts to
treat closed loops in
a grand canonical ensemble and corrects this with a
flux dependent chemical potential.
Here electro-chemical potentials characterize metallic
contacts.
For interacting systems it is in addition necessary
to distinguish the electrically isolated ring from
rings which interact via long range Coulomb forces
with other nearby conductors.
We use as a reference state a self-consistent
equilibrium state. The Coulomb interactions are determined
by the actual sample specific charge distribution
and not with respect to a flux averaged or ensemble
averaged charge density.

\section{Coulomb Driven Suppression of Level Hybridization}

Consider a one-dimensional loop with a weak disorder potential.
A stub, a wire of finite length,
is via a barrier coupled to this loop\cite{MBUNP}
(see Fig. 2).
This structure incorporates some features which
are typically encountered in multi-channel rings
with finite cross-sections\cite{CHEUN,BOUCH}:
As a function of flux a single particle energy exhibits
regions in which the state is nearly flux-insensitive $dE_{n}/d\Phi
\sim 0$ and behaves as if it were a
{\it localized} state. These localized regions are interrupted
by rapid changes as a function of flux and the state behaves as if
it were highly {\it mobile}.
We can view such a multichannel spectrum as a hybridization
of a highly mobile subsystem with a subsystem of localized states.
The simple example of Fig. 2  allows to investigate
the interplay of states which are localized in the stub with
highly mobile states in the loop.

We assume that the electro-static
potential can be taken uniform inside
the stub and can be taken uniform inside the loop.
Let $U_{s}$ denote the electro-static potential
(bottom of the conduction band) inside the
stub (index $s$). The electrostatic potential
inside the
ring (index $r$) is $U_{r}$.
If the barrier is not transparent the states in the stub
have energy
$E_{s,m}+eU_{s}$,
with $m = 1, 2, ..,M$.
The spectrum of the states in the ring is
$E_{r,n}(\Phi)+eU_{r}$
with $n = 1, 2, ..,N$.
A spectrum of such a system is shown in Fig. 3a.
The energies of the
flux sensitive (mobile) states of the loop are drawn as solid
lines. The energies of the
flux-insensitive (localized) states of the stub are shown
as broken lines.
The $N$ electrons in the loop give
a persistent current
\begin{eqnarray}
I_{N}(\Phi) = -c \sum_{n=1}^{N} dE_{r,n}(\Phi)/d\Phi
\label{eq05}
\end{eqnarray}
shown in Fig. 4.

Now let us make the barrier transparent.
Now the stub and the ring form one combined system.
Assume that the transparency of the barrier is very small.
The spectrum of the combined system will undergo only very small
changes where levels of the disconnected system intersect.
In a first step we evaluate the spectrum of the combined system
taking the potentials $U_{s}$ and $U_{r}$ to be known.
In a second step we will include the Coulomb interaction
to determine these potentials.
Consider any two levels $E_{s}$ and $E_{r}$ which
in the absence of transmission intersect.
In the combined system
the energy levels of these two states
have the hybridized energies
\begin{eqnarray}
E^{h}_{\pm} = (1/2)
(E_{s} +eU_{s} +
E_{r}+eU_{r} \pm \Delta )
\label{eq06}
\end{eqnarray}
where the energy gap between the two levels
is given by
\begin{eqnarray}
\Delta = \left((E_{s} +eU_{s} -
E_{r}-eU_{r})^{2} + 4 |t|^{2}\right)^{1/2}.
\label{eq07}
\end{eqnarray}
Here $|t|$ is the energy which couples the states of the
ring and the stub.
The spectrum $E^{h}_{l}$ with $l = 1,....,M+N$
of the combined system is shown in Fig. 3b.
It consists of pairs of states with eigenvalues
given by Eq. (\ref{eq06})
and consists of states $E^{h}_{l}=E_{s,m}+eU_{s}$
which do not intersect any level of the loop but
fall into a gap.
The persistent current of the hybridized system
(in the absence of Coulomb interactions)
is given by
$I^{h}(\Phi) = -c \sum_{l} dE^{h}_{l}/d\Phi$.
In the absence of interactions the flux dependence
still originates from the flux dependence of the
energies $E_{r,n}.$  Since the hybridized states occur,
except for the topmost occupied state, in pairs
the current is the sum of the persistent current
of the uncoupled loop with $N-1$ electrons plus the contribution
of the topmost occupied state with energy
$E^{h}_{-}$
in the hybridized system,
\begin{eqnarray}
I^{h}(\Phi) = I_{N-1}(\Phi)
-c dE^{h}_{-}/d\Phi.
\label{eq08}
\end{eqnarray}
For a small flux we have
$dE^{h}_{-}/d\Phi \sim dE_{N}/d\Phi$
and the persistent current of the combined system is the same
as that of the disconnected system.
But for a flux far beyond the hybridization point $\Phi_{*}$
we have
$dE^{h}_{-}/d\Phi \sim 0$. Thus beyond the hybridization
point the persistent current is that of an $N-1$ electron
loop. This behavior is shown in Fig. 4.
The current of $N-1$ electron states is typically
of opposite sign and smaller in magnitude then that
of the $N$ electron loop.
Thus the current of the hybridized system is typically
smaller in magnitude than the persistent current of the
decoupled system.
Below we will now show that if the Coulomb interactions
are taken into account and are of sufficient strength
the current of the combined system is in fact restored
to a value close to that of the uncoupled system.

The hybridized states
with energies $E^{h}_{\pm}$ are coherent
superpositions of a state that was originally localized in the stub
and a state that originally was confined to the ring.
As a consequence of the superposition the state
$E^{h}_{+}$
has
only a fractional charge $Q_{s+}$ inside the stub and has
a fractional charge $Q_{r+}$ inside the ring.
Similarly the state with energy $E^{h}_{-}$ has
a fractional charge $Q_{s-}$ inside the stub and has
a fractional charge $Q_{r-}$ inside the ring.
These charges can be found by differentiating the energies
of these states with respect to the potentials $U_{s}$
and $U_{r}$.
With $\alpha = r,s$ we find for the partial charges
\begin{eqnarray}
Q_{\alpha,\pm} = dE^{h}_{\pm}/dU_{\alpha}.
\label{eq09}
\end{eqnarray}
Of course the total charge in each filled state is
$Q_{s\pm}+Q_{r\pm} = e$.
{}From Eq. (\ref{eq09}) we see immediately that if both
states $\pm$ are filled then the combined charge of these
two states in the stub is
$Q_{s+}+Q_{s-}=e$ and in the ring is
$Q_{r+}+Q_{r-}=e$.
Consequently a net charge motion from the ring into the stub
occurs only if the topmost
occupied (empty)
state in the ring intersects an empty (occupied) state in the
stub as shown in Fig. 3b.
Thus the Coulomb interaction is entirely determined
by the charge motion in the topmost level $E^{h}_{-}.$

Before we proceed to evaluate the potentials we calculate
the variation of the charges of the topmost occupied level
$E^{h}_{-}$
in response to a small variation of the potentials.
There is a Lindhard function
$\Pi_{\alpha\beta},$
with $\alpha = r, s$ and $\beta = r,s$
which gives the charge response in the stub or in the ring
due to a variation of the potential in the stub or in the ring,
$dQ_{\alpha} = -\Pi_{\alpha\beta} dU_{\beta}.$
{}From Eq. (\ref{eq09}) we find
$\Pi_{\alpha\beta} =
d^{2}E^{h}_{-}/dU_{\alpha}dU_{\beta}$.
Using Eq. (\ref{eq06}) we find after a little algebra
$\Pi_{ss} =
\Pi_{rr} =
-\Pi_{rs} =
-\Pi_{sr} = - \Pi$
where
\begin{eqnarray}
\Pi = 2e^{2}|t|^{2}/\Delta^{3}
\label{eq10}
\end{eqnarray}
In the absence of interactions, i. e. for fixed potentials
$U_{s}$ and $U_{r}$,
the polarization function exhibits a sharp peek of magnitude
$1/|t|$ at the point of hybridization and is small
as soon as the difference in energies of the two states
exceeds $|t|^{2}$.
Next we want to characterize the variation in charge in the
stub and in the ring as a function of flux.
For fixed potentials an increment $d\Phi$ in the flux
causes a change in the charge on the stub given by
$(dQ_{s}/d\Phi)_{U} =
- (d(dE^{h}_{-}/dU_{s})/d\Phi)_{U}$.
Since at constant potential the flux dependence of the
energy stems from $E_{r,N}$ only we find
$(dQ_{s}/d\Phi)_{U} =
- (d^{2}E^{h}_{-}/dU_{s}dE_{r,N})(dE_{r,N}/d\Phi)$.
But
$ d^{2}E^{h}_{-}/dU_{s}dE_{r,N} =(1/e)
d^{2}E^{h}_{-}/dU_{s}dU_{r}$ and hence using
Eq. (\ref{eq10}) we find for the flux-induced charge
variation
\begin{eqnarray}
(dQ_{s}/d\Phi)_{U} = (1/e)
\Pi
(dE_{r,N}/d\Phi)
\label{eq11}
\end{eqnarray}
Since the total charge in each state is conserved
an increment in flux leads to a charge
$(dQ_{r}/d\Phi)_{U} = - (1/e)
\Pi
(dE_{r,N}/d\Phi)$ in the ring.
We are now ready to determine the self-consistent potentials.

An increment $d\Phi$
in the AB-flux changes the charge in the
stub by an amount
$(dQ_{r}/d\Phi)_{U} d\Phi.$
Screening of these charge causes a variation of the potentials
by $dU_{s}$ and $dU_{r}$. According to the Lindhard function
this causes an additional charge
variation in the stub given by
$-\Pi dU_{s}+ \Pi dU_{r}.$
The Coulomb interaction is taken into account with the
help of a geometrical capacitance $C$ between the stub and the
ring. Taking the total charge variation on the stub to be equal to
that permitted by the Coulomb interaction gives\cite{MBUNP}
\begin{eqnarray}
(dQ_{s}/d\Phi)_{U} d\Phi
-\Pi dU_{s}+ \Pi dU_{r} = C(dU_{s}-dU_{r}).
\label{eq12}
\end{eqnarray}
The charge in the ring obeys an equation which term by term
is identical to Eq. (\ref{eq12}) except for the sign of the
charges.
Thus Eq. (\ref{eq12}) determines only the difference in potentials.
The solution of Eq. (\ref{eq12}) is the difference
$v = v_{s} - v_{r}$ of the
characteristic potentials
$v_{s} =edU_{s}/d\Phi,$
$v_{r} =edU_{r}/d\Phi,$
\begin{eqnarray}
v = ed(U_{s}-U_{r})/d\Phi
= \Pi/(C+\Pi) (dE_{r,N}/d\Phi)
\label{eq13}
\end{eqnarray}
Integration of Eq. (\ref{eq13})
gives
\begin{eqnarray}
e(U_{s}- U_{r}) = \int^{\Phi} d\Phi^{\prime} \Pi/(C+\Pi)
(dE_{r,N}/d\Phi^{\prime}).
\label{eq14}
\end{eqnarray}
The Lindhard function $\Pi$ is also a function
of the difference of the potentials $e(U_{s}-U_{r})$.
Thus Eq. (\ref{eq16}) is similar to the self-consistent
equations encountered
in other problems, for instance the BCS gap equation.
We note that the Lindhard function modifies the geometrical
capacitance and gives rise to a quantum correction.
The charge induced in response to an external potential
is determined by an effective capacitance\cite{MBHTP,WEI94}
$1/C_{eff}=1/C+1/\Pi$.

Fortunately there is a simple limiting case in which
the solution of the self-consistent equation is obvious.
For a very small geometrical capacitance
the right hand side of Eq. (\ref{eq14}) is independent of the
potentials.
The characteristic potential is then
determined by the energy derivative of the topmost
state of the completely decoupled system,
$v(\Phi)= dE_{N}/d\Phi$.
Integration gives
\begin{eqnarray}
e(U_{s}- U_{r}) =
E_{r,N}(\Phi) - E_{r,N} (0)+ eU_{0}
\label{eq15}
\end{eqnarray}
where $U_{0}$ is the potential difference
between the conduction band bottoms of the ring and the
stub at zero flux.
Using Eq. (\ref{eq15}) in Eq. (\ref{eq07}) we find that for
the topmost occupied level
$\Delta
= ((E_{s}-
E_{r,N}(0)-eU_{0})^{2} + 4 |t|^{2})^{1/2}$ is {\it independent} of
flux.

Let us now show that in the limit of vanishing capacitance
the persistent current of the interacting system is
equal to that of the decoupled system.
At $kT = 0$
the total energy for the interacting system is
the sum of the kinetic energy (see Eq. (\ref{ekin}))
and the interaction energy\cite{KOHNW,NLANG}
$E_{c}$,
\begin{eqnarray}
E_{tot}(\Phi) = \sum_{n=1}^{N+M} E^{h}_{n}
-Q_{s}U_{s}-eQ_{r}U_{r}+ E_{c}.
\label{eq16}
\end{eqnarray}
Evaluating this energy we find
\begin{eqnarray}
E_{tot}(\Phi) = \sum_{m=1}^{M} E_{s,m}
+ \sum_{n=1}^{N-1} E_{r,n}(\Phi) + E^{h}_{-}(\Phi)-eU_{r}(\Phi)
-Q(U_{s}-U_{r})+ E_{c}.
\label{eq17}
\end{eqnarray}
But for very small capacitance
the charge imbalance $Q$ vanishes and the interaction
energy is a flux independent constant.
Furthermore, the difference of
the hybridized energy of the topmost state
and $eU_{r}(\Phi)$ is
$E^{h}(\Phi) -eU_{r}(\Phi) = E_{r,N}(\Phi)+ E_{0}$
with $E_{0}$ a flux independent energy.
Up to flux independent terms the total energy is that
of the completely decoupled ring.
Thus in the limit of small capacitance the
persistent current of the interacting
system is the same as that of the original, completely
decoupled system.
This is a consequence of the suppression of
level hybridization through Coulomb interaction.

Below we investigate the characteristic potential $v$ for
conductors which are open (see Fig. 1), i. e. connected
to electron-reservoirs.
This permits us to investigate the capacitances and
permits us to investigate the effect on the characteristic
potential of other nearby metallic bodies.

\section{Characteristic Potentials for an Open Normal Loop}

Consider the open conductors of Fig. 1.
We now relate the characteristic potentials for these
conductors
to electron
densities.
A variation in the electro-chemical potential
$\mu_{k}=E_{F,k}({\bf r})+eU({\bf r})$
by $d\mu_{k}$ can be accomplished in two ways:
We can either increase the Fermi energy by
$dE_{Fk}=d\mu_{k}$
or the electric potential by
$edU_{k}=d\mu_{k}.$
We imagine a two step process: in the first step
we increase the {\it chemical} potential in reservoir $k$ by
$d\mu_{k}$, keeping the electro-static potential fixed.
As a consequence, an additional charge density
$ (dn({\bf r}, k)/dE)_{U} d\mu_{k}$
is injected into the conductor.
In a second step we switch on the Coulomb interaction.
The added charges create an additional induced
electrical potential which in turn gives rise to
an induced charge distribution $dn_{ind}({\bf r})$.
Thus the total change in charge density due to an electro-chemical
potential variation in conductor k is
$dn_{k}({\bf r}) =
(dn({\bf r}, k)/dE)_{U} d\mu_{k} +
dn_{ind,k} ({\bf r}).$
Similarly, the variation of the electron density
due to a change in flux consists of two contributions:
We first evaluate the change in electron density
keeping the electric potential fixed.
In a second step we evaluate the potential due the
polarization of the sample caused by the increase in flux
and calculate the contribution of this potential due
to the variation in charge density
$dn_{k}({\bf r}) =
(dn({\bf r}, k)/d\Phi)_{U} d\Phi +
dn_{ind,k} ({\bf r}).$
In the presence of a variation of the electro-chemical potential
and a variation in flux we have thus
\begin{eqnarray}
dn_{k}({\bf r}) =
\left( \frac{dn_{k}({\bf r})}{dE}\right)_{U} d\mu_{k} +
\left(\frac{dn_{k}({\bf r})}{d\Phi}\right)_{U} d\Phi
+ dn_{ind,k}({\bf r}).
\label{eq20}
\end{eqnarray}
The induced density
$dn_{ind,k}({\bf r})$ in conductor k
generated by a variation in the electro-static potential
$dU ({\bf r})$
can be specified by the Lindhard function
(or polarization function)
$ \Pi_{k} ({\bf r}, {\bf r}^{\prime}),$
\begin{eqnarray}
dn_{ind,k} ({\bf r}) = - \int d^{3}r^{\prime}
\Pi_{k}({\bf r}, {\bf r}^{\prime}) e dU({\bf r}^{\prime})
\label{eq21}
\end{eqnarray}
with a potential given by Eq. (\ref{eq19}).
In Eq. (\ref{eq21}) the integral
$ d^{3}r^{\prime}$
can be taken over all
space enclosed by a volume which contains both
conductors including a portion of the reservoirs\cite{MBCAP}.
(This convention applies also
to subsequent volume integrals in this work).
To obtain an overall charge
neutral system the volume has to be chosen so large that
no electric field lines penetrate its surface.
Eq. (\ref{eq21}) is just the continuous
space analog of the Lindhard function specified by Eq. (\ref{eq11})
for the closed system.
Invariance of the charge distribution under simultaneous changes in
all electro-chemical potentials\cite{MBHTP,MBCAP}
implies that the integral over the first or
the second spatial argument of the Lindhard function is
equal to the density of states in conductor k,
\begin{eqnarray}
(dn_{k}({\bf r})/dE)_{U} = \int d^{3}r^{\prime}
\Pi_{k} ({\bf r}, {\bf r}^{\prime}).
\label{eq22}
\end{eqnarray}
The density response described by the Lindhard function
is a consequence of a change in the equilibrium potential.
Since an equilibrium density is an even function of magnetic flux,
the Lindhard function is also an even function
of the magnetic flux,
$\Pi_{k} (\Phi, {\bf r}^{\prime}, {\bf r})
=\Pi_{k} (-\Phi, {\bf r}^{\prime}, {\bf r})$.
This has the consequence that the electro-chemical capacitance
is an even function of the magnetic flux.
We emphasize that these symmetry properties are characteristic
for conductors connected to a single reservoir:
Reciprocity symmetries apply
for multiprobe conductors\cite{MBCAP,WEI94}.

In the insulator (index $ k=0$) separating the conductors, a
potential variation can polarize the insulator
and induce a charge density
$dn_{ind,0} ({\bf r}) = - \int d^{3}r^{\prime}
\Pi_{0} ({\bf r}, {\bf r}^{\prime}) e dU({\bf r}^{\prime})$.
No external charges reach the insulating region and thus instead
of Eq. (\ref{eq22}) we find that the volume integral of the
Lindhard function over either the first or the second argument vanishes,
$\int d^{3}r^{\prime}
\Pi_{0} ({\bf r}, {\bf r}^{\prime}) =
\int d^{3}r^{\prime}
\Pi_{0} ({\bf r}^{\prime}, {\bf r}) =0$.
Note that the Lindhard function which we introduced for the
closed loop (see Eq. (\ref{eq10})) also has this property.

Next we write down Poisson's equation for the potential $U$.
If we expand $U$ with respect to $d\mu_{k}$
we find that the characteristic function
$u_{k}$ is determined by\cite{MBCAP}
\begin{eqnarray}
-\Delta u_{k} ({\bf r}) + 4 \pi e^{2}
\int d^{3}r^{\prime} \sum_{l=0}^{l=3}
\Pi_{l} ({\bf r}, {\bf r}^{\prime})
u_{k} ({\bf r}^{\prime}) =
4 \pi e^{2} (dn_{k}({\bf r})/dE)_{U}.
\label{eq23}
\end{eqnarray}
Eq. (\ref{eq23}) contains
the sum of all Lindhard functions of all the conductors $l=1,2$
and of the insulating region $l=0$.
The density of states of conductor $k$ plays
the role of a source term for the characteristic function
$u_{k}$.
Similarly the characteristic function $v({\bf r})$ is a solution
of the Poisson equation
\begin{eqnarray}
-\Delta v({\bf r}) + 4 \pi e^{2}
\int d^{3}r^{\prime} \sum_{l=0}^{l=3}
\Pi_{l} ({\bf r}, {\bf r}^{\prime})
v({\bf r}^{\prime}) =
4 \pi e^{2} (dn_{1}({\bf r})/d\Phi)_{U}.
\label{eq24}
\end{eqnarray}
The flux-sensitivity of the
electron density in the conductor with the loop
is the source term of the characteristic function $v$.
If the source term is replaced
by a test charge
$ e\delta({\bf r}-{\bf r}_{0})$
which is concentrated at one point ${\bf r}_{0}$
the solution to Eq. (\ref{eq23}) and Eq. (\ref{eq24}) is
Green's function $g({\bf r}, {\bf r}_{0}).$
With the help of Green's function
we find for the characteristic function,
\begin{eqnarray}
u_{k} ({\bf r}) =
\int d^{3}r^{\prime} g ({\bf r}, {\bf r}^{\prime})
(dn_{k}({\bf r}^{\prime})/dE)_{U}.
\label{eq25}
\end{eqnarray}
Similarly the characteristic function $v({\bf r})$ is given by
\begin{eqnarray}
v({\bf r}) = \int d^{3}r^{\prime} g({\bf r}, {\bf r}^{\prime} )
(dn_{1}({\bf r}^{\prime})/d\Phi)_{U}.
\label{eq26}
\end{eqnarray}
Eq. (\ref{eq19}) implies for Green's function
the property\cite{MBCAP}
\begin{eqnarray}
\int d^{3}r^{\prime} g ({\bf r}, {\bf r}^{\prime})
\sum_{k} (dn_{k}({\bf r}^{\prime})/dE)_{U} = 1.
\label{eq27}
\end{eqnarray}
The same relationship follows
from the condition that the sum of all induced
charge densities
plus the test charge is zero.
Eq. (\ref{eq27}) will be used to demonstrate
charge and current conservation of the results derived below.

\section{Electro-Chemical and Flux-Induced Capacitance}

The characteristic potentials
derived above can now be used to find the electro-chemical
capacitance and a flux-induced capacitance.
Using
Eqs. (\ref{eq20}),
(\ref{eq21}),
and (\ref{eq25}),
the
total charge in conductor $k$
can be expressed in terms of density of states and Green's
function.
Differentiating the total
charge
$dQ_{k}$
with respect to the voltage
$dV_{l} = d\mu_{l}/e$
gives an electro-chemical capacitance\cite{MBCAP}
$C_{kl} =
edQ_{k}/d\mu_{l}$ given by
\begin{eqnarray}
C_{kl} = e^{2}
\int d^{3}r
\int d^{3}r^{\prime}
(dn_{k}({\bf r})/dE)_{U}
\left(\delta_{kl} \delta ({\bf r} - {\bf r}^{\prime})
- g ({\bf r}, {\bf r}^{\prime})
(dn_{l}({\bf r}^{\prime})/dE)\right)_{U}.
\label{eq28}
\end{eqnarray}
Eq. (\ref{eq28}) expresses the capacitances in terms of
the density of states of the reference state
and Green's function which mediates the Coulomb interactions.
Conservation of charge relates the capacitance coefficients
of our two-terminal conductor
as follows:
$C_{11}= C_{22} = -C_{12} = -C_{21}.$
To see this one makes use of
Eq. (\ref{eq27}).
Below we will use the abbreviation
$C_{\mu} = C_{11}$
where the index $\mu$ reminds us that we deal not with an electrostatic
capacitance but with an electro-chemical capacitance.

Next, in analogy to the electro-chemical
capacitance coefficients just discussed
we consider the flux induced capacitance
$C_{\Phi,k}= dQ_{k}/d\Phi.$ This is
the ratio of the piled up charge in conductor $k$ and the
increment in flux $\Phi$.
We find the flux-induced capacitances
\begin{eqnarray}
C_{\Phi,k} = e
\int d^{3}r
\int d^{3}r^{\prime}
\left(\delta_{1k} \delta ({\bf r} - {\bf r}^{\prime})
-(dn_{k}({\bf r})/dE)_{U}
g({\bf r}, {\bf r}^{\prime})\right)
(dn_{1}({\bf r}^{\prime})/d\Phi)_{U}.
\label{eq29}
\end{eqnarray}
These two coefficients are related.
Charge conservation implies that the sum of these two coefficients
is zero. This can again be demonstrated by using
Eq. (\ref{eq27}).
Hence we are left with one coefficient only and denote it by
$C_{\Phi}$,
$C_{\Phi} = C_{\Phi,1}= - C_{\Phi,2}$.
Whereas the electro-chemical
capacitance is an even (and periodic) function of flux
the flux-induced capacitance is an odd (and periodic) function of flux.

Consider now for a moment the conductor of Fig. 1a
which is in an electrically neutral environment.
Then according to Eq. (\ref{eq27})
the spatial integral over ${\bf r}$ of Green's function
and the density of states is just equal to 1.
Hence in this case both the capacitance $C_{\mu}$
and $C_{\Phi}$ vanish.

In the presence of time-dependent chemical potentials
and in the presence of a time-dependent oscillating flux
the current $I_{1}(t)$ measured at terminal $1$ is
determined by $I_{1}(t) = dQ_{1}/dt$ and the current
at terminal 2 (the gate) is determined by
$I_{2} = dQ_{2}/dt$.
Since charge is conserved, the current is conserved,
$I_{1}(t) + I_{2}(t) = 0$.
The current depends only on the difference of the
chemical potentials.
Taking the oscillating chemical potential difference to be
$d\mu_{1} - d\mu_{2} = edV_{\omega} exp(-i\omega t)$ and keeping
the flux fixed gives rise to a current
$dI_{\omega} = -i\omega C_{\mu} dV_{\omega}$.
Measurement of this current in a
zero-impedance external circuit determines the
electro-chemical capacitance.
An oscillatory flux with Fourier component
$d\Phi_{\omega}exp(-i\omega t)$ gives a current
$dI_{\omega} =
- i\omega C_{\Phi}d\Phi_{\omega} $
in a zero-impedance external circuit.

If the external circuit has a non-vanishing impedance an oscillatory
flux will also lead to an oscillating voltage.
In this case a total induced
current
\begin{eqnarray}
dI_{\omega} = -i\omega C_{\mu} dV_{\omega}
- i\omega C_{\Phi}d\Phi_{\omega}
\label{eq30}
\end{eqnarray}
is generated.
The voltage $dV_{\omega}$ depends on the external impedance.
For a circuit with impedance $Z(\omega) = - dV_{\omega}/dI_{\omega}$
an oscillatory flux generates a voltage
\begin{eqnarray}
dV_{\omega} =
- \frac{i\omega C_{\Phi}}{(1/Z_{\omega})-i\omega C_{\mu}}
d\Phi_{\omega} .
\label{eq31}
\end{eqnarray}
Since we deal with a capacitive effect it is useful
to measure the voltage capacitively and take the external impedance
to be capacitive
$Z(\omega) = 1/(-i\omega C_{ext}).$
In this case the measured voltage is
independent of frequency and given by
\begin{eqnarray}
dV_{\omega} =
\frac{C_{\Phi}}{C_{\mu}+C_{ext}}
d\Phi_{\omega}.
\label{eq32}
\end{eqnarray}
Our discussion of the conductor in Fig. 2 has shown that the
charge displaced by the flux is at best one electronic charge
and is less if interactions are taken into account.
If this is indicative also of the open conductor
in Fig. 1 then the magnitude of the flux-induced capacitance
is of the order of
$e/\Phi_{0}$  or
$C_{\Phi} = e^{2}/hc.$
This is equivalent to a "conductance" $G_{\Phi} = cC_{\Phi}= e^{2}/h$
equal to the fundamental conductance unit.
Screening will reduce these values but
both the measurement of the electro-chemical
capacitance and the flux-induced
capacitance should be feasible.

In this work we have discussed the response of the internal
electro-static potential of mesoscopic
conductors to small changes in an external parameter (flux,
electro-chemical potential) with the help of characteristic functions.
The characteristic potential $v$ is responsible for the enhancement
of the persistent current above a value achieved in the non-interacting
system. This potential also determines a flux-induced
capacitance. The characteristic function $u$ determines the electro-
chemical capacitance. Measurement of these
capacitances can, therefore, provide an experimental means
to determine the sensitivity of the electric potential.

\section{Note added in proof}
To obtain an enhancement of the persistent current in situations
different from that shown in Fig. 3 it is in general
necessary that the stub acts like an {\it acceptor} in the
hybridized system. The stub acts like an acceptor (rather
than a donor) if its density of states is much larger then that
of the loop. This corresponds to the situation in diffusive
multichannel
rings where we have many localized states and only a few
highly mobile states.

\section{Acknowledgement}
I would
like to acknowledge the support of the SERC and
the hospitality of the Cavendish Laboratory in Cambridge, U.K.,
where part of this work was done.
I also would like to acknowledge instructive discussions with
N. D. Lang which helped to correct an earlier version of this
manuscript.

\figure{(a) Ring with a lead connected to
an electron reservoir with an electro-chemical potential
$\mu_{1}$.
(b) The ring is separated by an
insulator from a
second conductor (gate) which is connected
to an electron-reservoir at an electro-chemical potential $\mu_{2}$.
\label{autonum}}

\figure{Normal electron ring which is via a tunneling
barrier with capacitance $C$ connected to a short wire (stub) of
finite length.
\label{autonum}}

\figure{Energy spectrum as a function of flux for the
conductor of Fig. 2. (a) The ring and the stub are completely
disconnected. (b) The barrier between the ring and the stub is
transparent. Coulomb interactions are not taken into account.
\label{autonum}}

\figure{Persistent current $I_{N}$ (solid line)
and $I_{N-1}$ (dashed line)
of the completely disconnected system
with $N$ and $N-1$ electrons in the ring.
For a transparent barrier and in the absence
of interactions the persistent current is
$I^{h}$ (dotted line).
$\Phi_{*}$ is the flux at which the topmost
energy level of the loop hybridizes with a localized state
of the stub.
In the presence of interactions the persistent current $I_{N}$
is recovered even for a transparent barrier.
\label{autonum}}

\end{document}